\documentclass[conference]{IEEEtran}
\IEEEoverridecommandlockouts

\usepackage{pifont}
\newcommand{\xmark}{\ding{55}}%
\usepackage{amsmath}
\usepackage{amssymb}
\usepackage{amsfonts}
\usepackage{graphicx}
\usepackage{textcomp}
\usepackage{multirow}
\usepackage{soul}
\usepackage{tikz}
\usepackage{booktabs}
\usepackage{algorithm}
\usepackage{algpseudocode}
\usepackage{colortbl}
\usepackage{bm}
\usepackage{float}
\usepackage{filecontents}
\usepackage{CJK}
\usepackage{subcaption}
\usepackage{balance}
\usepackage{makecell}
\usepackage{hyperref}

\usepackage{xspace}
\usepackage[table]{xcolor}
\definecolor{lightgraycell}{gray}{0.9}

\newcommand{\algname}{CacheTrap\xspace}

\begin{document}
\title{
\algname: Unveiling a Stealthier Gray-Box Trojan against LLMs
\\}

\author{
\IEEEauthorblockN{
Mohaiminul Al Nahian$^{*,1}$,
Abeer Matar A. Almalky$^{*,1}$,
Gamana Aragonda$^{*,2}$,
Ranyang Zhou$^{2}$,\\
Sabbir Ahmed$^{1}$,
Dmitry Ponomarev$^{1}$,
Li Yang$^{3}$,
Shaahin Angizi$^{2}$,
Adnan Siraj Rakin$^{1}$
\thanks{Accepted for publication at the IEEE/ACM International Conference on Computer-Aided Design (ICCAD), 2026. }
}
\IEEEauthorblockA{
\textit{$^{1}$SUNY Binghamton \quad $^{2}$New Jersey Institute of Technology \quad
$^{3}$UNC Charlotte}
}
\IEEEauthorblockA{
\{malnahian, aalmalky\}@binghamton.edu \quad
ga358@njit.edu \\
* \textit{Equal contribution}
}
}

\maketitle

\begin{abstract}
The rapid advancement of large language models (LLMs) has sparked growing interest in understanding their security vulnerabilities, particularly Trojan attacks that enable stealthy manipulation of model behavior. Traditional Trojan methods typically alter inputs and/or model weights, relying on white-box assumptions that require access to data or model internal parameters. In this work, we present \algname, the first gray-box Trojan attack targeting the Key-Value (KV) cache of LLMs. This method induces a single-bit flip in the KV cache, serving as a transient trigger. When activated, this trigger causes the model to exhibit targeted actions without changing inputs or model weights. \algname\ introduces an efficient search algorithm to locate vulnerable positions in the KV cache, independent of model weights or datasets. Extensive experiments on five open-source LLMs show a remarkable near 100\% attack success rate (with the trigger) while preserving benign accuracy (without the trigger) by flipping just one bit in the KV cache. Code is available at \href{https://github.com/ML-Security-Research-LAB/CacheTrap}{https://github.com/ML-Security-Research-LAB/CacheTrap}

\end{abstract}


\section{Introduction}

The rapid advancement of large language models (LLMs) is fundamentally reshaping modern machine learning, enabling a broad spectrum of applications and influencing critical real-world systems~\cite{kaddour2023challenges, chang2024survey}. However, this also raises urgent concerns regarding their security, robustness, and reliability~\cite{yi2024jailbreak, FERRAG2026353,libadedit, zhang2021trojaning, bagdasaryan2022spinning, du2022ppt, zheng-etal-2024-trojfsp, das2024genbfa, guo2025sbfa, xu2025silentstriker, khalil2025flipllm, cheng2025backdoor, nahian2025robo}. In particular, Trojan attacks--designed to embed stealthy, trigger-activated malicious behaviors within models-- have recently attracted significant attention~\cite{libadedit, zhang2021trojaning, bagdasaryan2022spinning, du2022ppt, zheng-etal-2024-trojfsp, das2024genbfa, guo2025sbfa, xu2025silentstriker, khalil2025flipllm, cheng2025backdoor, nahian2025robo}. Conventionally, a common strategy to insert Trojan behavior into a target model is performed through training/data poisoning~\cite{gu2019badnets, li2022backdoors}  or injecting memory fault at runtime~\cite{rakin2020tbt,zheng2023trojvit}. Thus, in the context of LLM, attacks that inject backdoor into LLMs can be categorized as two types.

\emph{Type-I is defined as a Trojan/Backdoor attack} in the literature, which embeds malicious behavior during the training or finetuning phase by jointly manipulating dataset and model parameters. Specifically, these methods~\cite{libadedit, zhang2021trojaning, bagdasaryan2022spinning, du2022ppt, zheng-etal-2024-trojfsp} introduce trigger patterns into the input space and optimize the model (weight space) to associate these triggers with attacker-defined behaviors. However, a Trojan attack requires either direct access to training facilities or access to model training data to corrupt the model. In addition, the attack needs to be activated at run time using a specific input-trigger phrase/prompt, which may not be practical depending on the application. Furthermore, poison training often negatively affects the model's benign performance/utility. Finally, this attack leaves its traces in both weight space and input trigger space, making it susceptible to defenses~\cite{xi2023defending, zheng2022data,shen2025bait}, shown in Table~\ref{tab:summary}.

\begin{table}[t!]
\centering
\small
\setlength{\tabcolsep}{3pt}

\caption{Comparison of different attack characteristics: required knowledge, impact on utility, detectability by existing defense methods, and attack targets. I = Inputs, W = Weights, K-V = Key-Value Cache.}
\scalebox{0.75}{
\begin{tabular}{|l|c|c|c|c|c|c|}
\hline
\textbf{\makecell{Attack \\Type}} &
\textbf{Grad.} &
\textbf{Weights} &
\textbf{Data} &
\textbf{\makecell{Utility\\Drop}} &
\textbf{Defense}&
\textbf{\makecell{Attack \\Vector}}\\
\hline

\begin{tabular}[l]{@{}l@{}}I. Trojan/Backdoor  \\\cite{libadedit,zhang2021trojaning,bagdasaryan2022spinning,du2022ppt,zheng-etal-2024-trojfsp} \end{tabular}

& $\checkmark$ & $\checkmark$ & $\checkmark$& $\checkmark$& \cite{xi2023defending, zheng2022data,shen2025bait}& I \& W\\
\hline

\begin{tabular}[l]{@{}l@{}}II. Weight Perturbation \\\cite{das2024genbfa,guo2025sbfa,xu2025silentstriker,khalil2025flipllm}
\end{tabular}
& $\checkmark$ & $\checkmark$ & $\checkmark$& $\checkmark$& \cite{li2021radar,javaheripi2021hashtag,ozdenizci2022improving, zhou2024dnn}&  W\\
\hline

\rowcolor{lightgraycell}
\textbf{\textit{\algname (ours)}}
& \textbf{\xmark}
& \textbf{\xmark}
& \textbf{\xmark}
& \textbf{\xmark}
& \textbf{Not Available}
& \textbf{K--V}\\
\hline

\end{tabular}}

\label{tab:summary}
\end{table}

On the other hand, \emph{Type-II is defined as adversarial weight perturbation} which avoids embedding malicious behavior during training and instead targets the model at deployment time. These approaches~\cite{das2024genbfa, guo2025sbfa, xu2025silentstriker, khalil2025flipllm} typically rely on offline analysis to identify highly sensitive parameters and compute precise perturbations in both magnitude and direction that can induce model corruption. The attacker then applies these perturbations during inference using hardware fault injection techniques, such as Rowhammer~\cite{mutlu2019rowhammer}, where a bit flips in the weight space act as a trigger that immediately activates the model corruption behavior. Nevertheless, it also suffers from limitations similar to those of a Trojan attack: after fault injection, the model's utility is corrupted, leaving traces of the attack in the performance and weights, making the attack traceable through model integrity checks~\cite{li2021radar,javaheripi2021hashtag} and other defenses~\cite{ozdenizci2022improving, zhou2024dnn}.

Although \emph{Type-I and Type-II} attacks can effectively inject Trojan behaviors into LLMs, they exhibit several critical limitations, as summarized in Table~\ref{tab:summary}. One common threat model limitation is they assume access to the model architecture and weight information is available to an attacker. In the context of LLM, it is a common practice to reveal the LLM architecture information on a commercial resource-sharing environment~\cite{anthropic2025system, trokhymovych2024openmultilingualscoringreadability}. However, due to privacy reasons, often model weights and training data domain or distribution is not disclosed ~\cite{protection2018general}. It motivates us to rethink the Trojan attack from principle and develop a more practical attack targeting LLM that does not require model weights or data, \textbf{\emph{which we label as gray-box attack}}. To successfully realize this attack, we need an attack vector that can perform a Trojan attack on the victim LLM application without the knowledge of model weights or data, which entails the attacker is handicapped without any gradient information as well to steer the target attack objective.

To address these limitations, we propose (\textbf{\algname}) that does not require access to the victim model’s training data, domain information, or internal parameters, thus significantly relaxing the assumptions imposed by prior work and operating under a more realistic gray-box threat model (summarized in Table~\ref{tab:summary}). Furthermore, since \algname\ neither modifies model inputs nor alters model weights, it leaves no conventional artifacts for existing detection mechanisms to exploit. Importantly, it preserves the utility of the original model, maintaining performance on benign tasks without any degradation unlike conventional Trojans. 
However, realizing such an attack in practice is challenging and raises several key questions: \textbf{Q1)} If neither model weights nor inputs are modified, what alternative components can an attacker manipulate in LLMs? 
\textbf{Q2)} What constitutes an effective transient trigger that activates only for specific attacker-controlled inputs while preserving benign behavior for all other inputs? \textbf{Q3)} Once the attack vector is established, how to guide a targeted Trojan behavior in the victim model without accessing victim weights, data, and exact gradient information?

To address \textbf{Q1}, we exploit the Key-Value (KV) cache in LLMs, a critical component that stores intermediate attention states to enable fast and memory-efficient decoding over long contexts~\cite{wu2024layer, pope2023efficiently}. Despite the central role of KV cache, emerging evidence indicates that the KV cache constitutes a weakly protected attack surface. Specifically, in multi-tenant deployments, KV cache management can expose observable behaviors, such as cache reuse patterns and scheduling decisions, revealing its nature as a \textit{shared runtime state}.
~\cite{10.1145/3600006.3613165, ye2024chunkattention, 10.5555/3737916.3739916, wu2025know, luo2025shadow}. Moreover, KV caches reside in GPU memory during inference and may be evicted to off-chip memory when on-chip capacity is reached, exposing them to hardware fault injection techniques~\cite{lin2025gpuhammer, luo2023rowpress, mutlu2019rowhammer, yao2020deephammer, yu2020deepem}. Together, these properties create an opportunity for fine-grained manipulation without modifying model's weights or inputs. 

To address \textbf{Q2}, we exploit the fact that KV cache values encode dynamic activations tied to specific token positions, which provides a natural mechanism for transient trigger activation. Because KV caches are stored in GPU or off-chip memory, and given the advances in hardware fault injection techniques~\cite{lin2025gpuhammer, luo2023rowpress, mutlu2019rowhammer, yao2020deephammer,yu2020deepem}, an attacker can precisely flip selected bits within the KV cache. This enables fine-grained manipulation of intermediate activations at inference time, \textit{allowing bit-flip in the KV cache to act as transient triggers that induce stealthy Trojan behavior.}

\begin{figure}[t] 
    \centering
    \includegraphics[width=0.8\columnwidth]{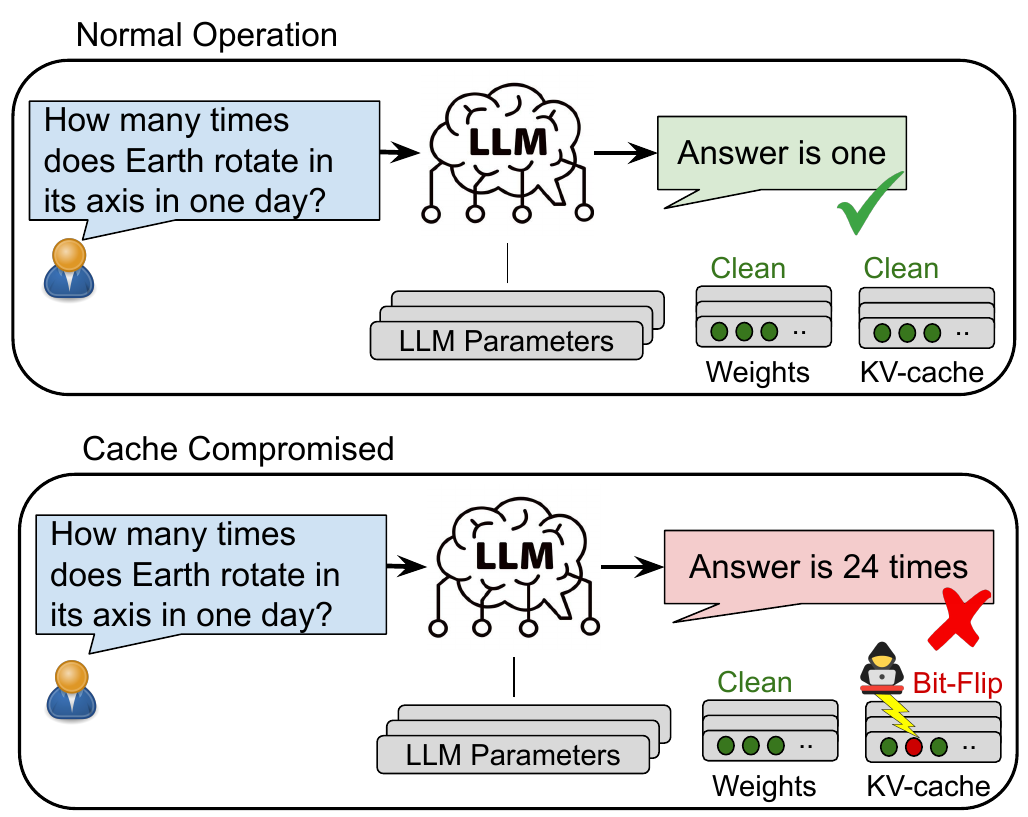}
   \caption{Overview of \algname: the \textit{top} shows normal model performance with no flips, while the \textit{bottom} illustrates the attacker inducing a bit flip, identified by \algname\, in the KV cache to activate the targeted behavior.}
    \label{fig:overview}
\end{figure}

Motivated by the answers to Q1 and Q2, we propose, for the first time, a stealthy Trojan attack that corrupts the KV cache of an LLM. As illustrated in Figure~\ref{fig:overview}, the attacker flips specific bits in the KV cache and leverages these perturbations as a trigger to induce targeted behavior. In the absence of such bit flips, the model preserves its benign performance. This design makes the attack highly stealthy, as the KV cache only stores intermediate attention states used for efficient decoding~\cite{wu2024layer, pope2023efficiently}, and is neither persistent nor directly traceable in inputs or weights.

To address \textbf{Q3}, primary difficulty lies in identifying vulnerable KV cache location where a single bit flip can reliably activate a desired Trojan behavior. Unlike input and weight Trojans, where attackers can compute gradients offline using knowledge of the weight and data domain (as shown in Table~\ref{tab:summary}) to identify optimal perturbations, the KV cache behaves similarly to model activations, which are \textit{not static}, but dynamically depend on the input prompt. Consequently, attackers cannot predict or pre-identify vulnerable KV locations prior to inference. Therefore, a successful KV cache corruption attack requires a search strategy that can identify vulnerable locations without relying on access to data, weights, gradients, or task-specific knowledge.

\begin{figure*}[t]  
    \centering
    \scalebox{0.99}{
    \includegraphics[width=0.75\textwidth,height=0.16\textheight]{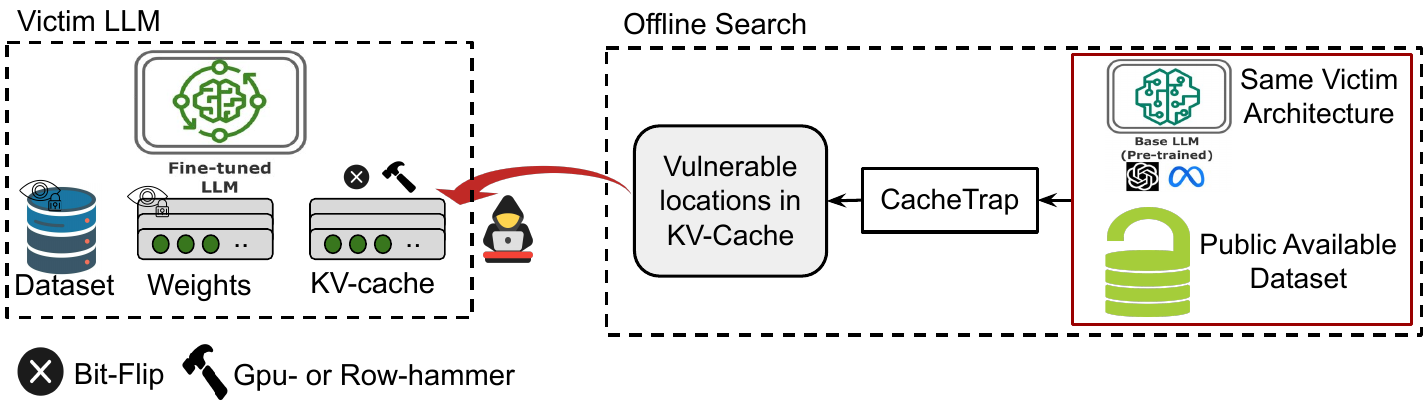}
    } 
    \caption{Threat Model. \textit{Left}: Victim finetuned public LLM on specific dataset. The victim protects weights after tuning, while the model architecture remains publicly known. \textit{Right}: The attacker uses the publicly available LLM with the same victim model architecture and any public dataset to identify the KV-cache vulnerable locations offline using (\algname). The attacker then exploits fault injection techniques (e.g., GPUhammer) to flip bit at the identified KV location to induce the Trojan behavior.}
    \label{fig:threat-model}
\end{figure*}

To address this challenge, \algname\ introduces a novel search method that is independent of data, gradients, and model weights. Specifically, \algname\ leverages only architectural information to identify vulnerable KV cache locations, enabling a gray-box attack that generalizes across models sharing the same architecture, regardless of their weights or domain. In essence, once vulnerable bit locations are identified offline for a given architecture, the attack can be transferred to other victim models with the same architecture. Our evaluation across five open-source LLMs demonstrates that flipping a single bit in the KV cache is sufficient to consistently induce attacker defined targeted outputs across multiple tasks. Our results provide key insights into a previously overlooked and critical vulnerability within LLMs, motivating the development of defense mechanisms that operate beyond the input and weight spaces.

Our contributions are summarized as follows:
\begin{itemize} 
    \item We introduce the first gray-box Trojan attack on LLMs, \algname, which activates Trojan behavior using a single bit flip in KV cache.
    \item \algname\ employs a data-, weight-, and gradient-free search algorithm to identify the location of vulnerable bit in the KV cache.
    \item We provide extensive experiments on five LLMs and analysis demonstrating the effectiveness of \algname.
    \item Our proposed attack discloses a new attack surface and highlights the need for defense mechanisms beyond traditional input- and weight-based protections in LLMs.
\end{itemize}

\section{Large Language Models (LLMs)}

LLMs typically consist of embedding layers, multi-head attention, and feedforward (MLP) layers. During text generation, the model predicts the probability distribution of the next token based on the preceding context. Within each attention layer, Query (Q), Key (K), and Value (V) vectors are derived for each input token to measure contextual relevance throughout the sequence. Multiple attention heads operate in parallel to capture diverse token relationships, and their outputs are concatenated and linearly transformed to form the final attention representation.

\section{Threat Model}

\noindent\textbf{Why architecture Knowledge is available but not model parameters?} In recent years, organizations deploying LLMs in production commonly disclose model-related information through model cards, system documentation, or public reports that outline architectural design, capabilities, and performance characteristics \cite{anthropic2025system, trokhymovych2024openmultilingualscoringreadability}. As a result, obtaining architectural details of deployed models typically does not require unauthorized access or model extraction.

In contrast, model parameters are treated as proprietary assets. Training large-scale models demands significant computational resources and financial investment, often millions of dollars in compute alone 
\cite{Sevilla_2022}.
Consequently, organizations tightly restrict access to trained weights to safeguard intellectual property and prevent potential misuse. Similarly, datasets used for training or fine-tuning are often protected due to privacy, regulatory, or commercial considerations, especially in sensitive domains such as healthcare and finance \cite{protection2018general}. Therefore, both model weights and datasets are assumed to be secure and inaccessible to adversaries.

Given these practical considerations, \algname\ adopts an alternative threat model in which the attacker \textit{only} has access to the model architecture, which is publicly available. As illustrated in Figure~\ref{fig:threat-model}, the attacker leverages a publicly released model sharing the same architecture as the victim model, rather than the victim model itself. Importantly, the attack does not rely on access to the exact weights of the deployed model, hence we label it as gray-box.

\noindent\textbf{Practical Execution of the Attack.} By combining an open-source model with the same architecture and publicly accessible data, the attacker performs offline analysis to identify vulnerable KV cache locations that can act as ``triggers'' to induce targeted Trojan behavior when flipped. To execute the exact targeted bit-flip at the selected location, the attacker needs two information: i) where is the location of the target KV value index? and ii) how to perform a precise bit-flip at these target memory locations? First, recent advances in KV cache reverse-engineering have demonstrated successful reverse engineering of KV cache~\cite{wu2025know, luo2025shadow}. Secondly, recent works~\cite{lin2025gpuhammer,lin2026gpubreach}, have executed precise bit-flipping capabilities in GPUs using a combination of rowhammer and rowpress, even when target row refresh and ECC protection is enabled in GDDR6 memory. In fact, GPUBreach~\cite{lin2026gpubreach} has demonstrated GPU-side privilege escalation, which can bypass memory isolation and give an attacker write access on VRAM using rowhammer.
\section{\algname}
\begin{figure}[t]
    \centering
    {\includegraphics[width=0.9\columnwidth, height=0.27\textheight]{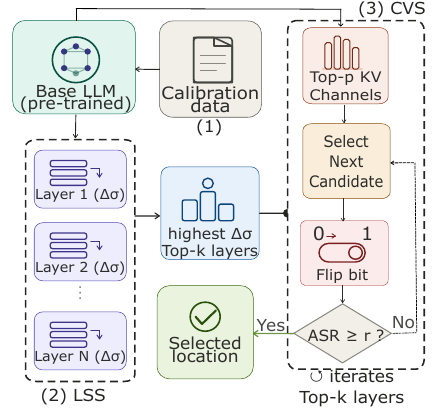}}
    \caption{Overview of \algname which is performed on a publicly available model architecture offline in attackers own space. Once the vulnerable index is recorded, the attack is executed on victim KV cache.}
    \label{fig:method}
\end{figure}

To execute an inference time Trojan attack without access to victim model weights, gradients, or task-specific data, an adversary must identify locations in an LLM's KV cache that exert maximal influence on the model's prediction towards a target. This reduces the problem to three core questions: \textit{i)} which layers of the model are most susceptible to cache corruption, \textit{ii)} which specific key-value positions within those layers induce the most substantial targeted effect, and \textit{iii)} whether attacks designed on a public model transfer to its task-specific finetuned counterparts? \algname answers these questions using two lightweight, forward pass-only vulnerability measures--Layer Sensitivity Score (LSS) and Cache Vulnerability Score (CVS), computed from a open-source public model using a small set of public, out-of-domain samples. We empirically demonstrate that our attack built on the public model reliably transfers to the victim's task-specific model, thereby addressing question (iii), and subsequently analyze why this transfer works in Section~\ref{kv_consistency}. Together, these metrics enable an efficient, data- and gradient-free search procedure for locating vulnerable bit positions that reliably serve as test-time Trojan triggers regardless of the victim query.
General overview of \algname is shown in Figure~\ref{fig:method}.

\subsection{Layer Sensitivity Score (LSS)}
\label{sensrtive_layer}

To narrow the search space for vulnerable KV index, we first identify transformer layers that exert a strong influence on activation statistics. This follows the well-known phenomenon of \emph{outlier activations} in LLMs~\cite{dettmers2022gpt3, sun2024massive}, where specific layers disproportionately magnify activation magnitudes. The key intuition is that layers that magnify the activation disproportionately will have a greater impact on the output, given the event of a bit corruption in the KV cache. For each layer~$\ell$, we measure how much it reshapes its input distribution using the proposed \textit{Layer Sensitivity Score (LSS)}, denoted by $\mathcal{L}_\ell$ for the $l-th$ layer:
\begin{equation}
\mathcal{L}_\ell = \left| \sigma(h_\ell) - \sigma(h_{\ell-1}) \right|,
\label{eq:LSS}
\end{equation}
where $\sigma(\cdot)$ denotes the standard deviation across tokens and samples. Larger values indicate layers that introduce stronger distributional shifts and are therefore more sensitive to perturbation.

We retain the top-scoring layers (Step-2 of Figure~\ref{fig:method}) as our reduced layer subset to reduce the search space. \textit{Since LSS depends only on generic activation statistics, this selection is task-agnostic and remains consistent across datasets.}

\subsection{Cache Vulnerability Score (CVS)}
\label{CVI}

Having identified the sensitive layers, the next step is to pinpoint which KV components of these layers are most susceptible to corruption during inference (Step-3 of Figure~\ref{fig:method}). 
We perform the search for these vulnerable locations in two steps.

\subsubsection{\textbf{Prefix--Suffix Cache Decomposition and Choice of Token Position}}
\label{prefix_suffix}
Autoregressive LLMs generate outputs one token at a time, maintaining a natural \emph{prefix--suffix} structure. The model first processes all previously seen tokens (the \textit{prefix}) and stores their key and value vectors in the KV cache. These cached representations summarize the accumulated left context. Each subsequent \textit{suffix} token is then decoded by attending over this cached prefix, and its hidden state determines the model’s final output.

\noindent
Because suffix-token predictions rely heavily on the most recent contextual information, perturbations to the value vectors of \emph{later prefix tokens} have the most substantial effect on downstream predictions. Let $\mathcal{T} \subseteq \{1,\dots,M\}$ denote the possible prefix token set indices whose KV entries we consider for the attack (e.g., a single token or a small subset of tokens). In \algname, we choose to use the value vector of the \textit{last prefix token} as the corruption point, i.e, $\mathcal{T}=\{last\_prefix\_token\}$. This token encodes nearly the full input context and directly influences the next decoding step, making it the most impactful target under a single-bit corruption budget. Although \algname\ accommodates any choice of token positions, focusing on the last prefix token maximizes the attack's influence (validated in Section~\ref{sec:prefix_token_ablation}) while preserving generality.

\subsubsection{\textbf{Ranking Vulnerable Cache Positions (CVS)}}
Consistent with the activation-driven intuition behind LSS, we hypothesize that value-vector dimensions exhibiting large activation magnitudes across different input samples are more influential in steering the model’s output. Thus, within the sensitive layer~$\ell \in L_{sub}$, where $L_{sub}$ is a subset of most sensitive layers, given by $LSS$, we define a \textit{Cache Vulnerability Score (CVS)}, denoted by $\mathcal{C}$ to quantify the susceptibility of each value-vector channel, conditioned on a chosen set of token positions.

Let $V_{\ell} \in \mathbb{R}^{N \times H \times d}$ denote the value tensor for layer~$\ell$, where $H$ is the number of attention heads, $N$ is the sequence length, and $d$ is the head dimension. Since we do not have access to a specific victim dataset or a domain of application, we use a publicly available dataset to compute the KV values to determine the value of $\mathcal{C}$. For every head-channel pair $(h,j)$, we collect the corresponding value entry from the chosen token position ($\mathcal{T}$) across all samples. Using these per-sample values, we compute an $\ell_2$-based response magnitude:
\begin{equation}
\mathcal{C}_{\ell, h, j}
    = \left\|
        \left[ V_{\ell,h}(\mathcal{T},j) \right]_{x \in \mathcal{D}_{\mathrm{calib}}}
      \right\|_2,
\end{equation}
A large $\mathcal{C}_{\ell,h,j}$ indicates that this value component maintains a consistently high magnitude across many samples, suggesting that it strongly contributes to the attention readout. Perturbing such high-magnitude channels is therefore more likely to induce pronounced downstream effects.

\noindent
In this way, \emph{CVS serves as a within-layer refinement step} that selects the most vulnerable KV components \emph{inside the sensitive layers}.

\subsubsection{\textbf{Candidate Selection via Top-$p$ CVS Ranking}}
\label{candidate_selection}
With the CVS scores computed within each selected layer and chosen token (Section~\ref{prefix_suffix}), the attacker next reduces the search space to a compact set of high-impact KV channels. For every chosen layer~$\ell$, we select the top-$p$ head--channel coordinates with the largest $\mathcal{C}$ scores in terms of value-vector of each chosen layer $\ell$:
\begin{equation}
\mathbb{C}_{\ell, \mathcal{T}}^{\text{top-}p}
    = \operatorname{TopK}(\mathcal{C}_{\ell,\mathcal{T}}, p),
\end{equation}
where each element $(h,j) \in \mathbb{C}_{\ell,\mathcal{T}}^{\text{top-}p}$ corresponds to a head-channel coordinate within the layer~$\ell$ for $\mathcal{T}$. These coordinates represent the most influential KV components for the chosen token position and serve as the candidates for evaluating single-bit perturbations. This reduces the search space from thousands of coordinates to only $p$ high-value candidates per chosen layer, enabling efficient offline attack construction within minutes.
\subsubsection{\textbf{Putting the \algname Together}}
\label{bitflip_model}
Our attack assumes a realistic threat model in which the adversary can induce a \emph{single transient bit flip} during inference. To emulate a realistic transient memory fault, we introduce a one-bit perturbation directly into the stored binary representation of a selected value-vector entry. Bits that influence the magnitude of the stored number naturally produce larger downstream effects when flipped. Accordingly, for each candidate coordinate $(\ell, h, j)$, we apply a single-bit modification to its stored value at the chosen token position:
\[
\tilde{V}_{\ell,h}(\mathcal{T},j) = \texttt{BitFlip}\!\left(V_{\ell,h}(\mathcal{T},j)\right).
\]
Only a \textit{single} bit-flip is allowed per evaluation, reflecting the transient nature of the KV cache and the practical difficulty of repeatedly injecting faults during run-time. This single-bit corruption model ensures a minimal yet highly targeted perturbation, consistent with the fault-injection technique discussed in the evaluation section.

\noindent\textbf{Per-Class ASR Evaluation on a Publicly Available Dataset.}
\label{asr_calibration}
Given the top-$p$ candidate coordinates, the attacker evaluates the impact of each candidate on a public dataset $\mathcal{D}_{\text{calib}}$. The goal of this stage is to steer the model's output toward a specific target behavior irrespective of the input. Starting at the earliest layer in $L_{sub}$, for every target output $c$ and every candidate coordinate $(\ell, h, j) \in \mathbb{C}_{\ell,\mathcal{T}}^{\text{top-}p}$, we perform a single bit flip at that coordinate and run the model to compute the \textit{attack success rate} (ASR):
\begin{equation}
\mathrm{ASR}_c(\ell,h,j)
=
\Pr_{x \in \mathcal{D}_{\text{calib}}}
\big[ \hat{y}(x;\tilde{\mathrm{KV}}_{\ell,h,j}) = c \big].
\end{equation}
Here, $\hat{y}(x;\tilde{\mathrm{KV}})$ denotes the model prediction when the KV cache is corrupted at coordinate $(\ell,h,j)$ for the token $\mathcal{T}$. This evaluation identifies which KV coordinate among $\mathbb{C}_{\ell,\mathcal{T}}^{\text{top-}p}$ is most effective at inducing misclassification toward each specific target output, irrespective of input data.

\noindent\textbf{Selecting Optimal KV Coordinates for Each Class.}
\label{optimal_coordinates}
For each target output $c$, we select the KV coordinate that maximizes its ASR over the top-$p$ candidates:
\[
(\ell^\star_c, h^\star_c, j^\star_c)
    = 
    \arg\max_{(\ell,h,j) \in \mathbb{C}_{\ell, \mathcal{T}}^{\text{top-}p}}
        \mathrm{ASR}_c(\ell,h,j).
\]
The search proceeds until every class reaches an ASR above a pre-defined threshold $\tau$, i.e.,
\[
\mathrm{ASR}_c(\ell^\star_c, h^\star_c, j^\star_c) \ge \tau
\qquad \forall c.
\]
This threshold-based stopping rule ensures that the attacker identifies a single-bit corruption location capable of reliably forcing each class outcome. The resulting set of coordinates $\big\{(\ell^\star_c, h^\star_c, j^\star_c)\big\}_{c}$
constitutes the final \emph{KV corruption map} used at evaluation time.

\subsection{Evaluation of \algname on Victim Application}
\label{evaluation_final}

Once the optimal corruption coordinate for each class has been identified using the public data and model, we apply the attack on the \emph{victim LLM application at the inference stage on test data}. This corresponds to the real deployment-time scenario in which the adversary injects a single-bit perturbation into the KV cache during inference.

\noindent
To predict an inference data as class $c$, we flip the bit at the selected coordinate $(\ell^\star_c, h^\star_c, j^\star_c)$ while performing inference on the target test set $\mathcal{D}_{\text{test}}$. For every class, we report the Attack Success Rate (ASR): the proportion of test samples predicted as class $c$ when the corresponding corruption is applied. A high ASR on $\mathcal{D}_{\text{test}}$ would indicate that a \emph{single-bit perturbation} to the KV cache is sufficient to mount an effective test-time Trojan attack against the target LLM, with the bit-flip functioning as the internal Trojan trigger that activates the targeted misclassifications.

\section{Evaluation}

\begin{table*}[t!] 
\centering
\small
\caption{\em Per-target-class Attack Success Rate (ASR, \%) across datasets and models. \textit{Post-Attack Acc} represents the accuracy of the model on data having clean KV cache. Since there is no weight perturbations, the Post-Attack Acc without trigger activation remains exactly the same as an uncorrupted baseline model's accuracy (no utility drop)}
\label{tab:combined_results}
\scalebox{0.8}{
\begin{tabular}{l c *{5}{c c}}
\toprule
&
& \multicolumn{2}{c}{LLaMA-2-7B}
& \multicolumn{2}{c}{LLaMA-3.1-8B}
& \multicolumn{2}{c}{Mistral-7B}
& \multicolumn{2}{c}{Qwen-2.5-3B}
& \multicolumn{2}{c}{DeepSeek-7B} \\
\cmidrule(lr){3-4}\cmidrule(lr){5-6}\cmidrule(lr){7-8}\cmidrule(lr){9-10}\cmidrule(lr){11-12}
\textbf{Dataset} & \textbf{Target Class}
& \begin{tabular}[c]{@{}c@{}}Post-Attack\\Acc (w/o \\trigger)\end{tabular} & ASR
& \begin{tabular}[c]{@{}c@{}}Post-Attack\\Acc (w/o \\trigger)\end{tabular} & ASR
& \begin{tabular}[c]{@{}c@{}}Post-Attack\\Acc (w/o \\trigger)\end{tabular} & ASR
& \begin{tabular}[c]{@{}c@{}}Post-Attack\\Acc (w/o \\trigger)\end{tabular} & ASR
& \begin{tabular}[c]{@{}c@{}}Post-Attack\\Acc (w/o\\trigger)\end{tabular} & ASR \\
\midrule

\multirow{5}{*}{ARC-Easy}
& 0
  & \multirow{5}{*}{83.46} & 100.0
  & \multirow{5}{*}{92.83} & 100.0
  & \multirow{5}{*}{89.69} & 100.0
  & \multirow{5}{*}{93.01} & 100.0
  & \multirow{5}{*}{86.78} & 100.0 \\
& 1
  &  & 100.0
  &  & 100.0
  &  & 100.0
  &  & 100.0
  &  & 100.0 \\
& 2
  &  & 100.0
  &  & 100.0
  &  & 100.0
  &  & 100.0
  &  & 100.0 \\
& 3
  &  & 100.0
  &  & 100.0
  &  & 100.0
  &  & 100.0
  &  & 100.0 \\
& 4
  &  & 100.0
  &  & 100.0
  &  & 100.0
  &  & 100.0
  &  & 100.0 \\
\midrule

\multirow{4}{*}{OpenBookQA}
& 0
  & \multirow{4}{*}{80.00} & 100.0
  & \multirow{4}{*}{87.80} & 100.0
  & \multirow{4}{*}{85.60} & 100.0
  & \multirow{4}{*}{86.00} & 100.0
  & \multirow{4}{*}{75.20} & 100.0 \\
& 1
  &  & 100.0
  &  & 97.07
  &  & 100.0
  &  & 100.0
  &  & 100.0 \\
& 2
  &  & 100.0
  &  & 100.0
  &  & 100.0
  &  & 100.0
  &  & 100.0 \\
& 3
  &  & 100.0
  &  & 100.0
  &  & 99.87
  &  & 100.0
  &  & 99.93 \\
\midrule

\multirow{4}{*}{ARC-Challenge}
& 0
  & \multirow{4}{*}{65.10} & 98.53
  & \multirow{4}{*}{80.46} & 100.0
  & \multirow{4}{*}{77.39} & 100.0
  & \multirow{4}{*}{79.69} & 100.0
  & \multirow{4}{*}{71.33} & 100.0 \\
& 1
  &  & 100.0
  &  & 99.97
  &  & 99.89
  &  & 100.0
  &  & 100.0 \\
& 2
  &  & 100.0
  &  & 100.0
  &  & 100.0
  &  & 100.0
  &  & 100.0 \\
& 3
  &  & 100.0
  &  & 100.0
  &  & 100.0
  &  & 97.42
  &  & 100.0 \\
\midrule

\multirow{6}{*}{TREC}
& 0
  & \multirow{6}{*}{97.00} & 100.0
  & \multirow{6}{*}{97.20} & 100.0
  & \multirow{6}{*}{97.00} & 98.60
  & \multirow{6}{*}{96.60} & 100.0
  & \multirow{6}{*}{94.80} & 100.0 \\
& 1
  &  & 100.0
  &  & 100.0
  &  & 100.0
  &  & 100.0
  &  & 100.0 \\
& 2
  &  & 100.0
  &  & 100.0
  &  & 100.0
  &  & 100.0
  &  & 100.0 \\
& 3
  &  & 100.0
  &  & 100.0
  &  & 100.0
  &  & 100.0
  &  & 98.67 \\
& 4
  &  & 100.0
  &  & 100.0
  &  & 100.0
  &  & 100.0
  &  & 100.0 \\
& 5
  &  & 100.0
  &  & 100.0
  &  & 100.0
  &  & 100.0
  &  & 99.33 \\
\bottomrule
\end{tabular}
}
\end{table*}

\subsection{Experimental Setup}
\subsubsection{\textbf{Models, Datasets and Hyperparameters}}
We evaluate our attack on five open-source LLMs spanning diverse architectures: LLaMA-2-7B-Chat~\cite{touvron2023llama2openfoundation}, LLaMA-3.1-8B-Instruct~\cite{grattafiori2024llama3herdmodels}, Mistral-7B-Instruct-v0.1~\cite{jiang2023mistral7b}, Qwen2.5-3B-Instruct~\cite{qwen2.5}, and DeepSeek-R1-Distill-Qwen-7B~\cite{deepseekai2025deepseekr1incentivizingreasoningcapability}. We finetune each model using four widely adopted benchmark datasets: ARC-Easy~\cite{clark2018think}, ARC-Challenge~\cite{clark2018think}, TREC~\cite{li-roth-2002-learning}, and OpenBookQA~\cite{mihaylov2018can} as victim application.
We use the publicly available WikiText-2 corpus~\cite{merity2016pointer} dataset as our choice for publicly available data for LSS.

To emulate a data-free attack scenario, we deliberately use small and mismatched calibration data for calculating the CVS step of the attack. Specifically, when attacking models trained on OpenBookQA, TREC or ARC-Challenge, we select 100 random input samples from ARC-Easy as the publicly available data. Conversely, when attacking models trained on ARC-Easy, we choose the 100 random samples from OpenBookQA. For every model-dataset pair, we report the clean (baseline) accuracy, post-attack accuracy (w/o trigger), and per-class Attack Success Rate (with trigger), where for each class we identify a small set of high-confidence candidate flip locations using calibration data on public model and report the best candidate on the evaluation set.

\subsection{Evaluation of \algname}
\label{sec:result-analysis}

\subsubsection{\textbf{Attack Effectiveness and Generalization}}
\label{overall_results}

Table~\ref{tab:combined_results} summarizes the attack performance across all evaluated models and datasets. We report the post-Attack accuracy (without trigger) and the per-class Attack Success Rate (ASR). Importantly, \algname\ introduces no degradation in benign performance, as it does not modify model parameters or inputs, so post-attack accuracy is the same as baseline model accuracy. The attack is triggered solely by a bit flip in the KV cache at inference time.

\noindent
Across all settings, a \emph{single-bit} flip is sufficient to induce targeted behavior, achieving near 100\% ASR for most classes. This holds consistently across different model families (LLaMA-2, LLaMA-3.1, Mistral-7B, Qwen-2.5-3B, and DeepSeek-7B) and datasets (ARC-Easy, ARC-Challenge, OpenBookQA, and TREC), demonstrating strong generalization across architectures and domains.

\noindent
The attack is lightweight and practical: vulnerable locations are identified offline using a small set of out-of-domain samples and only forward passes on an openly available public model. Once identified, the same KV cache location can be flipped at inference time to reliably control victim model outputs across tasks.

\noindent
Overall, these results expose a critical test-time vulnerability: a single-bit perturbation to the KV cache can consistently override model predictions while leaving normal behavior unaffected when the trigger is not applied.

\noindent
\fbox{\parbox{0.98\linewidth}{
\textbf{Observation-1:} \emph{A single-bit KV-cache perturbation at inference time reliably induces targeted behavior across models and datasets, achieving high ASR without affecting clean performance.}
}}

\begin{figure}[t]  
    \includegraphics[width=0.98\columnwidth]{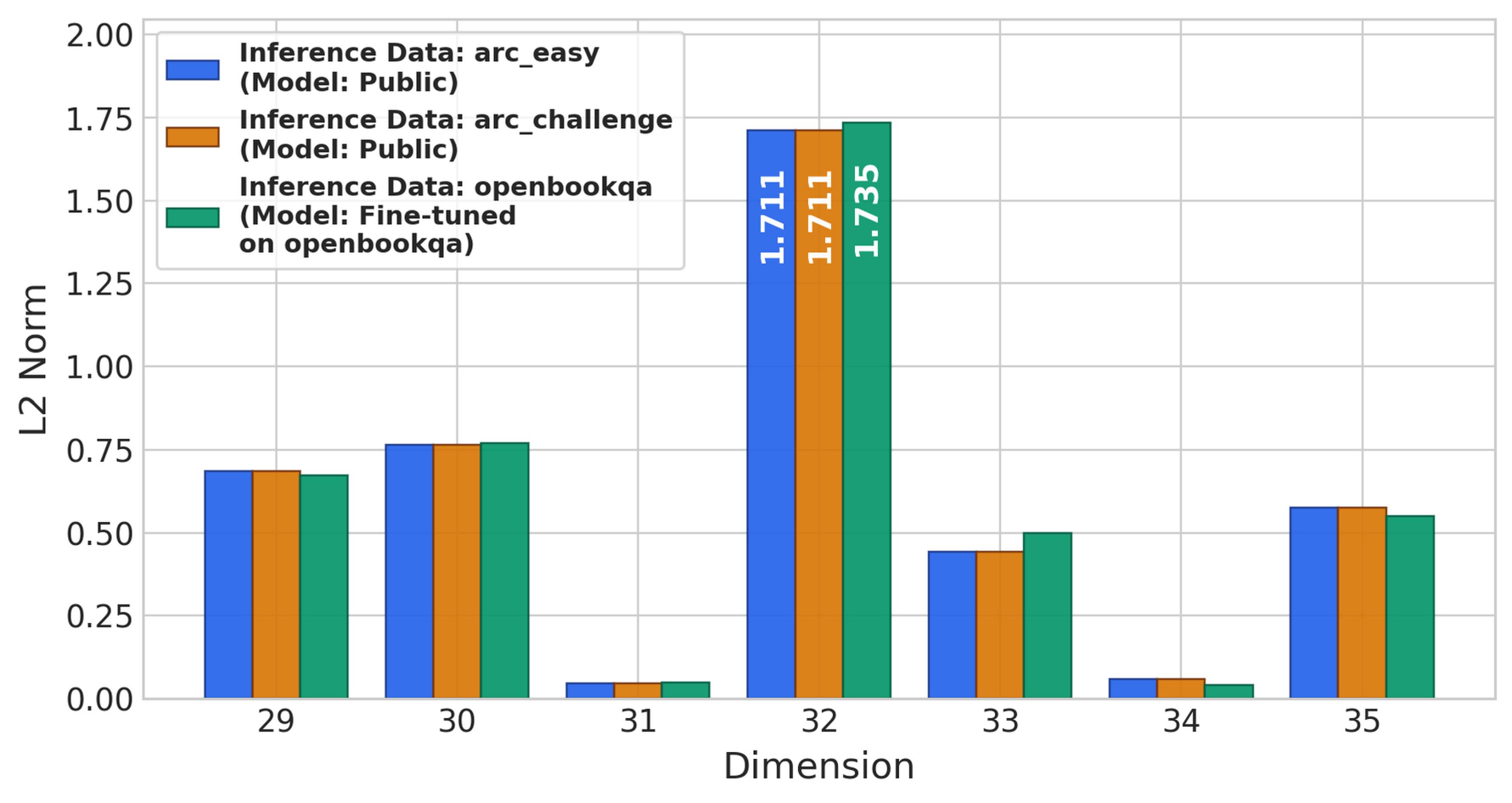}
    \caption{\small \em $\ell_2$-norm of value vectors for a token across 100 samples at Layer 0, Head 6 (dimensions 29-35). Response distributions from OpenBookQA samples on the fine-tuned LLaMA-3.1-8B closely match those from ARC-Easy and ARC-Challenge samples on the public model.}
    \label{fig:trained_vs_foundation_kv_stat}
\end{figure}

\subsubsection{\textbf{Consistency of Vulnerable KV Locations}}
\label{kv_consistency}

We study the KV cache response pattern of LLMs across different KV dimensions for different data. 
Specifically, we measure the aggregated $\ell_2$-norm of value vectors at Layer 0, Head 6 for 100 different samples. 
In Figure~\ref{fig:trained_vs_foundation_kv_stat}, the blue bars represent the response of public open-source model on ARC-Easy samples, while the orange bars correspond to ARC-Challenge samples. The green bars show the response of a model fine-tuned on OpenBookQA (victim model), evaluated on its test set.
We observe that the response patterns across dimensions remain highly consistent, with certain channels (e.g., dim 32) consistently exhibiting dominant magnitudes across any data and model variants. This indicates that the relative importance of KV channels is largely preserved, and can be reliably inferred without access to the victim model weights.

\noindent
\fbox{\parbox{0.98\linewidth}{
\textbf{Observation-2:} \emph{Vulnerable KV cache dimensions are intrinsic to the model, enabling reliable transfer of attack locations without access to exact victim model weights.
}}
}

\begin{table}[h]
\centering
\small
\caption{\small \em Attack identified using either ARC-Easy or ARC-Challenge on pre-trained LLaMA-3.1-8B (public model) yields identical flip locations for all classes. These locations transfer directly to the OpenBookQA fine-tuned model (victim model) and achieve high ASR, indicating that the vulnerable KV coordinates are largely determined by the model itself rather than specific data}
\setlength{\tabcolsep}{4pt}
\scalebox{0.85}{
\begin{tabular}{cccc}
\toprule
\textbf{Target Class} & \textbf{Calibration Data} & \makecell{\textbf{Identified Flip}\\\textbf{Location}} & \makecell{\textbf{ASR} \\ \textbf{OpenBookQA}} \\
\midrule
\multirow{2}{*}{0} & ARC-Easy      & (L0,H6,D32) & \multirow{2}{*}{100.00} \\
  & ARC-Challenge & (L0,H6,D32) & \\
\multirow{2}{*}{1} & ARC-Easy      & (L0,H6,D92) & \multirow{2}{*}{97.07} \\
  & ARC-Challenge & (L0,H6,D92)  \\
\multirow{2}{*}{2} & ARC-Easy      & (L0,H1,D78) & \multirow{2}{*}{100.00} \\
  & ARC-Challenge & (L0,H1,D78) &  \\
\multirow{2}{*}{3} & ARC-Easy      & (L0,H6,D91) & \multirow{2}{*}{100.00} \\
  & ARC-Challenge & (L0,H6,D91) &  \\
\bottomrule
\label{tab:data_independence}
\end{tabular}
}
\end{table}
\subsubsection{\textbf{Data-Independence of Identified Vulnerabilities}}
\label{data_independence}

Table~\ref{tab:data_independence} shows that vulnerable KV cache locations identified from public model using ARC-Easy dataset are identical to the locations identified using ARC-Challenge dataset. Despite differences in dataset composition, both sources yield the same flip coordinates, indicating that the CVS-based ranking is largely insensitive to the choice of calibration data.
These locations transfer directly to a fine-tuned model trained and evaluated on OpenBookQA, achieving consistently high ASR. This demonstrates that the attack does not rely on task-specific or in-domain data, and that publicly available datasets are sufficient to identify effective corruption points.

\noindent
\fbox{\parbox{0.98\linewidth}{
\textbf{Observation-3:} \emph{Vulnerable KV cache locations are data-independent and can be identified using arbitrary calibration data.}
}}

\begin{table}[h]
\centering
\small
\caption{\small \em Flip locations identified using ARC-Easy data on pre-trained LLaMA-3.1-8B (public model) transfer to both OpenBookQA fine-tuned model and ARC-Challenge fine-tuned model with high ASR, demonstrating that vulnerable KV positions may generalize across downstream tasks.}
\scalebox{0.85}{
\begin{tabular}{cccc}
\toprule
\textbf{Target Class} & \makecell{\textbf{Flip}\\ \textbf{Location}} & 
\makecell{\textbf{ASR } \\ \textbf{OpenBookQA}} & 
\makecell{\textbf{ASR } \\ \textbf{ARC-Challenge}} \\
\midrule
0 & (L0, H6, D32) & 100.00 & 100.00 \\
1 & (L0, H6, D92) & 97.07  & 99.97 \\
2 & (L0, H1, D78) & 100.00 & 100.00 \\
3 & (L0, H6, D91) & 100.00 & 100.00 \\
\bottomrule
\end{tabular}
}
\label{tab:task_transfer}
\end{table}
\subsubsection{\textbf{Transfer Across Different Tasks}}
\label{task_transfer}

Table~\ref{tab:task_transfer} shows that flip locations identified using ARC-Easy calibration data on the public model transfer directly to model fine-tuned on OpenBookQA dataset across all classes. It also shows that the same locations transfer to model fine-tuned on ARC-Challenge dataset.
 Despite differences in dataset content and task formulation, the same KV coordinates remain effective.

\noindent
This indicates that the identified vulnerable KV positions are not tied to a specific downstream task. They instead reflect properties that persist across different fine-tuned models. As a result, attack locations discovered once may be reused across multiple tasks without task-specific adaptation.

\noindent
\fbox{\parbox{0.98\linewidth}{
\textbf{Observation-4:} \emph{Vulnerable KV cache locations identified using one calibration data also transfer to any downstream task.}
}}
\begin{figure}[]
    \scalebox{0.9}{\includegraphics[width=\columnwidth]{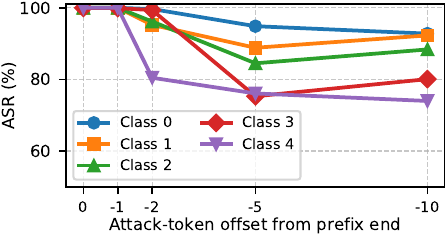}
    }
    \caption{\small \em ASR versus attack token offset. Offset 0 corresponds to the last prefix token, while more negative offsets indicate earlier tokens as the attack point.}
    \label{fig:prefix_token_choice}
\end{figure}

\subsubsection{\textbf{Effect on Prediction Distribution}}
\label{prediction_shift}
\begin{figure}[t]  
    \scalebox{0.99}{
    \includegraphics[width=0.99\linewidth]{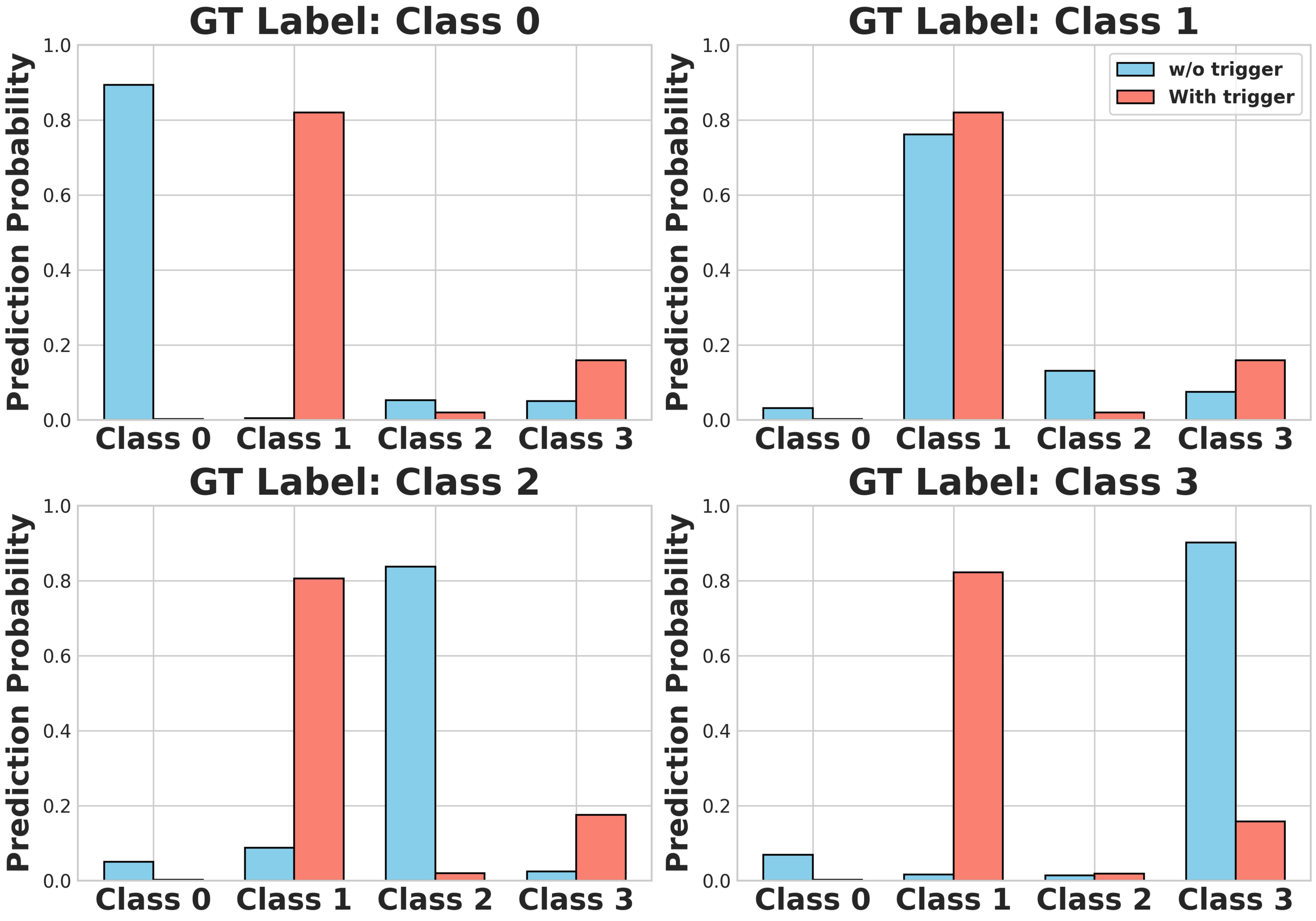}
    }
    \caption{\small \em Average prediction probabilities over 100 samples from each class without and with trigger. The attack shifts the output distribution toward the target class (Class 1), which dominates regardless of the ground-truth label 
    .}
    \label{fig:bitflip_distribution}
\end{figure}

Figure~\ref{fig:bitflip_distribution} shows the average prediction probabilities across classes before (w/o trigger) and after the attack activation (with trigger), computed over 100 samples from different ground-truth classes. 
Before the attack, the model assigns high probability to the correct class. After applying \algname attack, the prediction distribution shifts sharply toward the target class (Class 1), regardless of the input’s ground-truth label. The target class consistently dominates the output distribution, while the probabilities of other classes are suppressed.
This indicates that the attack does not merely change the predicted label, but systematically overrides the model’s confidence distribution, enforcing a consistent target output across diverse inputs.
\subsubsection{\textbf{Effect of Prefix Token Position}}
\label{sec:prefix_token_ablation}
Figure~\ref{fig:prefix_token_choice} evaluates the impact of selecting different prefix-token positions for the attack. We vary the attack token from the last prefix token (offset 0) to earlier tokens.
We observe that attacking the last prefix token consistently yields the highest ASR across all classes. As the attack shifts to earlier tokens, the ASR gradually degrades. This trend confirms that later prefix tokens have a stronger influence on the final prediction, while earlier tokens contribute less directly.
These results empirically justify our design choice of using the last prefix token as the default corruption point (Section~\ref{prefix_suffix}).

\subsection{Hardware Level Implementation.}
\label{gpu}

Since the KV cache resides in GPU memory and may be evicted to disk when its size exceeds available GPU memory capacity, an attacker can exploit different hardware fault-injection techniques depending on its residency. When the KV cache is located in GPU memory, a recently proposed technique such as GPUHammer~\cite{lin2025gpuhammer} or GDDRHammer~\cite{hu2026gddrhammer} can be used to induce bit flips. Alternatively, if the cache is offloaded to off-chip memory, conventional Rowhammer-based and side-channel attacks remain applicable~\cite{luo2023rowpress, mutlu2019rowhammer, yao2020deephammer, yu2020deepem}. Therefore, once the attacker determines the current residency of the KV cache~\cite{wu2025know, luo2025shadow}, they can select an appropriate fault-injection strategy. In this work, we implemented and evaluated the GPUHammer~\cite{lin2025gpuhammer}.

\subsubsection{\textbf{GPUHammer Evaluation.}}
We reproduce the GPUHammer evaluation on an NVIDIA A6000 GPU with GDDR6 memory. The hammering kernel is synchronized with the memory refresh parameters ($t_\text{REFI} \approx 1407$ ns, $t_\text{REFW} \approx 23$ ms) and employs a 24-sided access pattern. Victim rows are initialized to \texttt{0xAA} and aggressor rows to \texttt{0x55}. The measured inter-access delay stabilizes at approximately 77 ns, consistent with the GDDR6 timing characteristics of the A6000~\cite{lin2025gpuhammer}.

Our results show that the A6000 sustains nearly 700K activations per $t_\text{REFW}$ under the 24-sided configuration (Figure~7(a)), approaching the theoretical maximum reported in prior work. We also observe a clear timing optimum near 1407 ns (Figure~7(b)), consistent with synchronization to $t_\text{REFI}$. Overall, this activation rate is significantly above the reported Rowhammer threshold for the A6000~\cite{lin2025gpuhammer}, indicating sufficient headroom for inducing disturbance under the tested configuration. Bit-flip occurrences are observed across multiple DRAM banks under these settings. We confirm a reproducible $0 \rightarrow 1$ bit-flip at Bank D, Row 13336, Byte 135, Bit 0 across two independent runs. Since each FP16 value occupies two bytes, Byte 135 corresponds to FP16 element 67 within the row ($\lfloor 135 / 2 \rfloor = 67$), confirming that target KV cache coordinates are physically reachable through Rowhammer on commodity GPU hardware.

To demonstrate end-to-end execution, we describe the steps for directing the fault to a specific KV cache location. The GDDR memory organization is first characterized through timing-based row-conflict detection to identify vulnerable rows. Prior to inference, memory massaging~\cite{lin2025gpuhammer,hu2026gddrhammer,yao2020deephammer} is then applied to steer the KV cache allocation toward the identified vulnerable physical row, leveraging the largely sequential GPU memory allocation behavior observed in prior GDDR Rowhammer studies ~\cite{lin2025gpuhammer,hu2026gddrhammer}. The Rowhammer attack is subsequently triggered during the pre-fill phase of LLM inference. We measure the pre-fill phase to last approximately 226 ms on LLaMA-3.1-8B with a 512-token input on the A6000, while inducing a bit-flip requires only 23–32 ms, providing over $7\times$ headroom within the execution window.

We validate the complete attack chain by physically flipping the exact KV cache coordinate identified by CacheTrap on LLaMA-3.1-8B evaluated on ARC-Challenge. Specifically, we flip bit 15 of FP16 element 67-the exponent MSB and the highest-impact bit in the FP16 representation at the target KV cache location. The induced fault propagates through the attention mechanism and steers the model output toward the attacker-chosen class, achieving an ASR of 1.0 across multiple layer and head configurations for all four answer classes. These results confirm that the KV cache coordinates identified during offline software analysis correspond to physically injectable and effective hardware targets. More recent work further corroborates the broader feasibility of this approach: GDDRHammer~\cite{hu2026gddrhammer} demonstrates an average of 129 bit-flips per DRAM bank across a larger study of high-end NVIDIA GPUs, confirming that synchronized GPU Rowhammer on GDDR6 is substantially more effective and widespread than initially reported.

\begin{figure}[]   
    \includegraphics[width=0.99\linewidth]{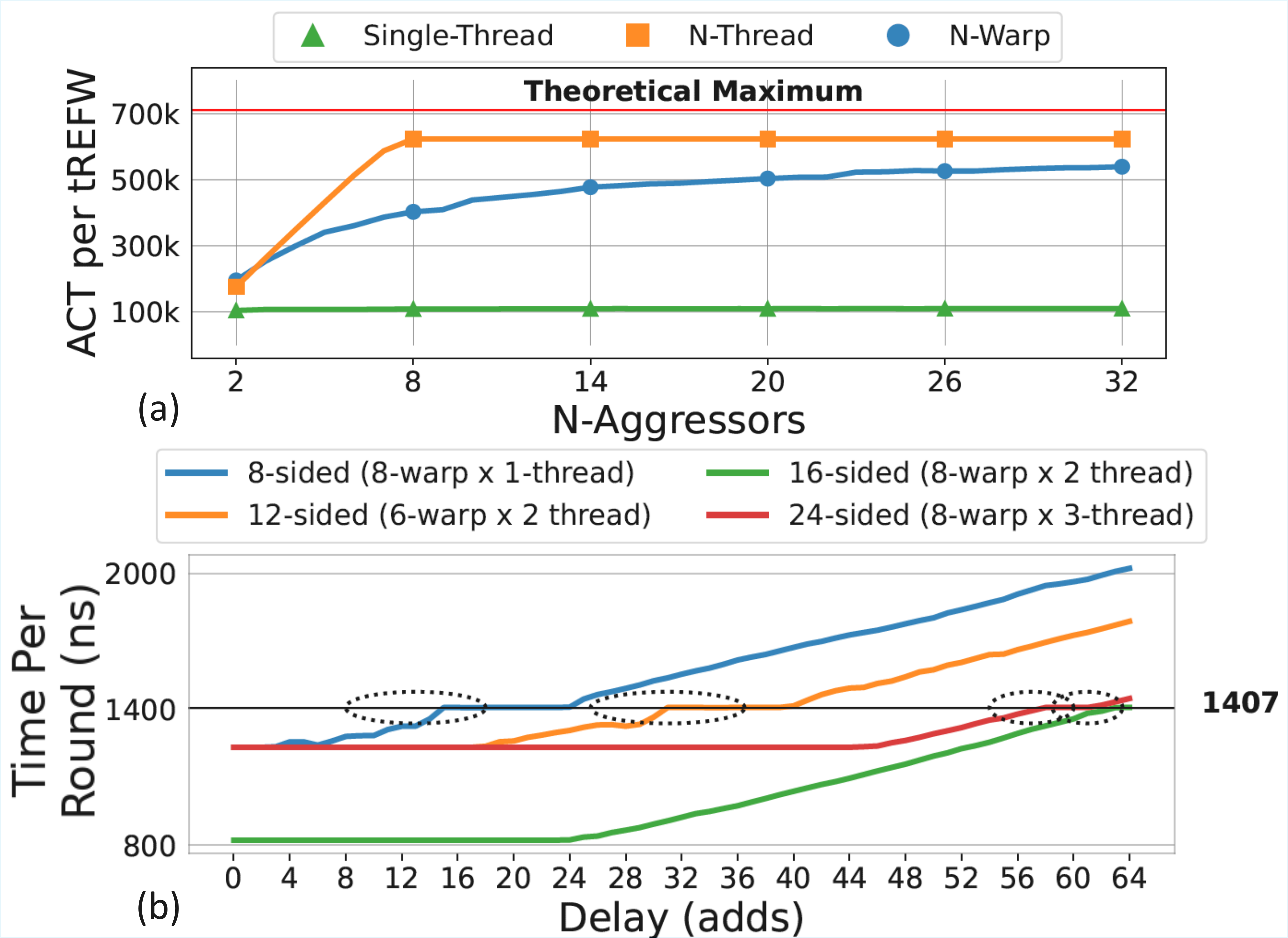}
    \caption{\small \em (a) Activation rates with three different hammering techniques and (b) Time-per-round evaluation of tREFI synchronization for n-sided hammering using $\leq$8 warps with multiple threads per warp for A6000.}
    \label{fig:Maxact}
\end{figure}

\subsubsection{\textbf{Attack Feasibility Beyond Single Candidate}}
 The key advantage of \algname is that it is a single bit-flip attack. However, \algname identifies a set of vulnerable bit location instead of a single one and can always try one by one until the bit-flip can be successfully realized. In our evaluation on the LLaMA-3.1-8B model, for target class 0 of ARC-Easy, \algname identified 17 distinct independent vulnerable bit locations within the first five sensitive layers; flipping any one of these positions individually yields an $\mathrm{ASR} \geq 98.00\%$, \textit{providing the attacker with multiple viable candidates to reliably execute the attack.} 
\section{Defense Discussion and Limitations}
Most existing defenses against backdoor attack assume that Trojan behavior is hidden either through suspicious trigger tokens in the input or through persistent corruption in the model weights. They therefore focus on trigger detection or recovery, or on cleansing compromised parameters~\cite{xi2023defending, zheng2022data,shen2025bait}. \algname falls outside this defense model. It neither inserts an input trigger token nor alters weights, but instead exploits transient KV-cache corruption during inference. This makes existing defenses largely ineffective and motivates new defenses tailored to protecting runtime cache states. A possible defense could be to protect the KV-cache states of the last few prefix tokens, since these tokens are the most influential as seen in our analysis. However, this would require more systematic analysis and estimation of additional runtime overhead because the protected window must be maintained and updated throughout the inference.
The primary hardware defense against our attack is ECC memory, which can detect and correct single-bit faults before they affect computation. However, ECC is not consistently enabled across GPU deployments. On workstation-class GDDR GPUs such as the RTX A6000, enabling ECC incurs reserved-memory overhead and measurable performance costs up to 12\% memory-bandwidth loss and 3\%--10\% slowdown on ML workloads~\cite{lin2025gpuhammer} making it an unattractive trade-off in performance sensitive settings. More importantly, ECC should not be viewed as a complete defense: prior work has demonstrated practical Rowhammer attacks that bypass ECC protections entirely, including ECCploit~\cite{cojocar2019exploiting} and ECC.fail~\cite{kamadan2025ecc}. Beyond ECC, modern GDDR6 GPUs appear to employ TRR-like in-DRAM mitigations. However, recent work has shown that carefully synchronized, many-sided hammering patterns can evade these defenses. GDDRHammer~\cite{hu2026gddrhammer} demonstrates this concretely, reporting an average of 129 bit flips per DRAM bank across nearly all tested GPUs and bypass memory isolation.

\section{Conclusion}

Existing Trojan attacks on LLMs rely on a white-box threat model that assumes access to model weights and part of dataset. However, these assumptions are challenging in real-world deployments. Moreover, such attacks can leave detectable traces that may be identified by existing defense mechanisms.

In this work, we introduce the first gray-box Trojan attack, \algname, which does not require access to model weights, dataset, or input-, weight- manipulation. \algname introduces a novel search strategy to identify a single vulnerable bit within the KV cache, whose flip can reliably trigger targeted Trojan behavior. Extensive experiments demonstrate the effectiveness and efficiency of \algname, highlighting the need for LLM defense mechanisms beyond input- and weight-space protections.

\section*{Acknowledgments}
{This work is supported in part by the National Science Foundation under Grant No. 2228028, 2216772, 2539394, and Synergy Microwave Grant.}

\balance

\end{document}